# Spatial Measures of Socioeconomic Deprivation:
## An Application to Four Midwestern Industrial Cities

Scott W. Hegerty
Department of Economics
Northeastern Illinois University
Chicago, IL 60625
S-Hegerty@neiu.edu

## ABSTRACT

Decades of economic decline have led to areas of increased deprivation in a number of U.S. inner cities, which can be linked to adverse health and other outcomes. Yet the calculation of a single "deprivation" index, which has received wide application in Britain and elsewhere in the world, involves a choice of variables and methods that have not been directly compared in the American context. This study creates four related measures—using two sets of variables and two weighting schemes—to create such indices for block groups in Buffalo, Cleveland, Detroit, and Milwaukee. After examining the indices' similarities, we then map concentrations of high deprivation in each city and analyze their relationships to income, racial makeup, and transportation usage. Overall, we find certain measures to have higher correlations than others, but that all show deprivation to be linked with lower incomes and a higher nonwhite population.

JEL Classification: R23, C10

Keywords: Economic Deprivation, United States, Spatial Methods

# I. Introduction

The process of economic restructuring, begun decades ago in the United States, has had enormous consequences for the country's industrial regions. In particular, the "Rust Belt" cities of the Northeast and Midwest have undergone depopulation, disinvestment, and reduced median incomes; social consequences such as higher crime and reduced life expectancy are also common in these inner cities. Studying the spatial patterns of these cities' socioeconomic conditions is paramount if appropriate policies are to be implemented to alleviate these problems.

One spatial measure of these conditions is a "deprivation index," which captures a number of economic and social indicators into a single variable. While these have been applied worldwide, relatively few studies have examined U.S. inner cities specifically. In addition, although there are many options of variables to include and mathematical approaches to construct an index, there are few side-by-side comparisons. This study addresses both issues simultaneously, calculating deprivation indices for four industrial U.S. cities, with two combinations of variables and two weighting schemes. After comparing the four measures, and finding them to broadly similar statistically, we examine their spatial distributions within each city. We also calculate their correlations with a variety of other socioeconomic variables, finding highly deprived areas to be poorer and less white, and to use more public transportation, than other areas of each city.

Much of the literature regarding deprivation has originated outside the United States, with a focus on antipoverty programs public health. In one seminal study, Townsend (1987) argues that existing measures introduced during the 1970s and 1980s misrepresent the distribution and severity of deprivation. He then measures disadvantage in Britain using 77 indicators including diet, clothing, housing, education, and integration and social inclusion.

This measure can be used as a benchmark against which other indices can be compared. Morris and Carstairs (1991) find that out of five separate measures, their own and Townsend's perform best when examining correlations with various health measures. Carstairs (1995) performs a review of earlier publications, noting the variety of deprivation measures used. The author then finds high correlations between mortality and health statistics with various deprivation indices and other socioeconomic variables.

Many other studies of this type also focus on areas outside the United States. Salmond *et al.* (1998) conduct principal component analysis (PCA) to capture the common variance in a set of ten variables in New Zealand. These include the receipt of benefits, the unemployment rate, schooling, occupancy, as well as housing tenure (renting versus owning), car access, the presence of single-parent families, and whether householders are separated or divorced. The authors then find correlations with cancer and mortality rates. Salmond and Crampton (2002) extend this type of analysis for small geographic areas.

Other studies have expanded these measures to make them more appropriate to areas outside urban Britain, often finding deprivation beyond this limited area. Langlois and Kitchen (2001), use PCA to construct a deprivation index for Montreal in the year 1996, finding deprivation outside the inner city. Pacione (2002) examines rural Scotland, noting that many existing indices that focus on urban areas in Britain include characteristics such as marital status and professional occupations. The author's PCA-based index, including more rural characteristics, includes self-employment and region-specific-sector employment, as well as (Gaelic) language knowledge. Bell *et al.* (2007) look at Vancouver using Geographic Information Systems (GIS), and allow for subjectivity and expert opinions in addition to purely statistical data in constructing a multivariable index. Seaman *et al.* (2015) construct a Scottish variant of the British Index of Multiple Deprivation (IMD) using it to explain excess mortality in Glasgow and other cities.

Outside the highly developed world, measures often include an entirely new set of variables. Noble *et al.* (2010), examine small areas in South Africa (3,799 wards), and Noble and Wright (2013) create an index for an even smaller unit of measurement. This index, based on the 2001 South African IMD, include not only the "standard" characteristics such as education, employment, and poverty, but also whether residents lack a TV or radio, flush toilets, water, or telephone, or live in a shack. Yu *et al.* (2011) include location quotients (LQs) for unemployment and other economic variables in Guangzhou, China, as well as LQs for the proportions of retirees and percentages of households without a toilet or other amenities. Studying Andalusia, Spain, Rodero-Cosano *et al.* (2014) use multiple variables to capture education, employment, infrastructure, income, health, and housing, as well as the lack of telephone or air conditioning, and the presence of crime, noise, and pollution. Vasquez *et al.* (2016), examining deprivation among immigrants in Chile from 1992 to 2012, include disability status alongside employment and education. While Broadway and Jesty (1998) examine Canadian inner cities, relatively few studies focus on U.S. urban areas. Smith (2009) examine environmental inequality in Detroit and Portland (Oregon). Using PCA to construct a deprivation measure, they do not find it to be closely linked to the location of Superfund (Federal hazardous waste storage) sites, which tend to be located in black neighborhoods rather than specifically poor ones.

Regardless of their specific method of construction, these measures are subject to a number of additional criticisms. Deas and Robson (2003) note the shortcomings of Britain's IMD, particularly with regard to large urban areas. In particular, data might be unreliable, and large areas might require more services. In addition, crime, air quality, and other measures of environmental degradation are excluded from the IMD. Bertin *et al.* (2014) also question whether an urban-based measure can be applied to rural areas. Perhaps most importantly, in a critical review of the literature, Fu *et al.* (2015) note that census-based methods include age and gender norms, as well as bias regarding housing type. It is this type of criticism that informs our study, leading us to ignore such variables, mentioned above, as the prevalence of single-parent households, car ownership, or housing tenure type.

The diversity of variables and methods used to calculate deprivation indices, as well as the relative paucity of studies that use them in American contexts, form the basis of the current study. To that end, we calculate a variety of related deprivation indices for the U.S. cities of Buffalo, Cleveland, Detroit, and Milwaukee. All are fairly large, regionally significant industrial cities that have undergone various degrees of decline and economic transition since the 1960s. After comparing our measures, we examine their spatial distribution and their correlations with other socioeconomic variables. This paper proceeds as follows. Section II describes our methodology. Section III details the results. Section IV concludes.

## II. Methodology

We use income and other socioeconomic data at the block-group level for the year 2014, from the United States Census Bureau, for this study. The cities of Buffalo, Cleveland, Detroit, and Milwaukee are chosen because they share certain similarities (They are large "Rust Belt" Midwestern cities with population below 1 million, so that "global cities" such as Chicago are excluded). All analyses are conducted in ArcGIS 10.3 or GeoDa, with the principal components calculated using Eviews.

We first note certain spatial patterns, particularly regarding income and other types of inequality within each city. Figure 1 depicts median income for our four cities. It is lowest on the East Sides of Buffalo and Cleveland, as well as the near North Side of Milwaukee. Detroit's income is not confined to any "inner city" area, because nearly all of the city except the edges (located near suburbs) are low-income. Figure 2 shows each city's relative proportions of nonwhite residents; these spatial patterns are similar to the income map. Nearly all of Detroit is majority nonwhite; additional areas (with relatively high income but relatively few whites) include the West Side of Buffalo and the far Northwest side of Milwaukee.

It is these spatial patterns of economic and other social characteristics that we seek to examine quantitatively. This study aims to calculate and compare various measures of economic deprivation, and then having done so, to examine their relationships with income, racial, and other socioeconomic characteristics of each city. Here, we combine a set of socioeconomic variables for each city's block groups into a single "deprivation" index. Our main decisions are how to assign the weights by which each component enters the index, and which variables to include. We begin by choosing between a narrower (four-variable) and a broader (six-variable) set of variables to include, and for each, then choose two alternative calculation methods.

As noted above, many indices of deprivation in the literature often are unsuitable for a developed country such as the United States, or they make value judgements about family structure or other living arrangements. We make no effort to include the vast number of components that were used in Townsend's (1987) original analysis. Keeping this in mind, we focus mainly on income, poverty, and education when choosing our variables. To that end, our four-variable "narrow" measure includes: the percentage of households in poverty, the percentage of workers older than 16 who are unemployed, the percentage of people over age 25 without a high-school diploma, and the percentage of households that receive nutrition assistance (SNAP). The broader definition also adds the vacancy rate and the percentage by which each block group's median income deviates from the citywide median. These latter two are entered separately because they might fail to capture true deprivation. For instance, an "urban prairie" would have few units to be vacant, and a poor city's median income is low to begin with. Nonetheless, we are able to judge whether "simpler is better," or if additional variables do indeed increase our index's explanatory power. Summary statistics for each component variable are presented in Table 1. As might be expected, all four cities are poor; unemployment is much higher than the national average, and highest in Detroit.

Our next choice comes in assigning weights. Here, we compare two methods: variance-based weights and principal components analysis (PCA). Variance-based weights deflate each component by each series' own standard deviation, as follows:

$$SD4 = \frac{percpov}{\sigma_{pp}} + \frac{unemp}{\sigma_{u}} + \frac{noHS}{\sigma_{HS}} + \frac{percSNAP}{\sigma_{pS}} \tag{1a}$$

$$SD6 = \frac{percpov}{\sigma_{pp}} + \frac{unemp}{\sigma_{u}} + \frac{noHS}{\sigma_{HS}} + \frac{percSNAP}{\sigma_{pS}} + \frac{percvac}{\sigma_{pvac}} - \frac{percMedY}{\sigma_{Y}} \tag{1b}$$

This is primarily dome so that the component with the most variance does not dominate the index.[1] Weights are calculated separately for each city, so it is possible that this could introduce some sort of bias. We therefore next apply PCA, which extracts the combination of multiple variables into the principal component that captures the maximum variation among the variables. Weights are depicted as factor loadings; we find that each city has exactly one principal component for both the 4- and 6-variable measures, called *PCA4* and *PCA6*, respectively.

Having created a total of 16 deprivation measures—four each for four cities, we next apply a variety of spatial and non-spatial statistical tools. First, we examine spatial autocorrelation for each of our deprivation indices. We use Moran's *I* as the standard measure of this association. This statistic calculates deviations from the mean both for a variable within each block group and for the block group's neighbors:

$$I = \frac{n}{\sum_{i=1}^{n}\sum_{j=1}^{n} w_{ij}} \frac{\sum_{i=1}^{n}\sum_{j=1}^{n} w_{ij}(x_i - \bar{x})(x_j - \bar{x})}{\sum_{i=1}^{n}(x_i - \bar{x})^2} \qquad (2).$$

---

[1] See Hegerty (2013), who compares standard-deviation and PCA weighting schemes to examine indexes in the foreign exchange market.

Much like traditional autocorrelation (usually used in a time-series context), higher values indicate that nearby areas affect one another. While there are a variety of choices for the weights matrix *w*, which defines what exactly is a "neighbor," we follow LeSage (2014) and choose a simple, first-order Queen continuity matrix. Here, only directly adjacent block groups (even those that only touch at one point) are given the value of 1, and others are given a weight of zero.[2]

Next, we calculate correlations among our four deprivation measure for each city, to address whether the calculation measures produce similar results. These are not spatially weighted in any way. Not only do we calculate the standard (parametric) Pearson *ρ* statistic, we also consider two nonparametric measures: Spearman's *ρ* and Kendall's *τ*. Nonparametric measures do not assume that data are normally distributed, and, because they use ranks to calculate their statistics, they are therefore less sensitive to outliers. We see below that the indices are highly correlated with one another.

We then look for deprivation "clusters" using the Getis-Ord (1995) *G\** statistic. This measure calculates "hot spots" and "cold spots," where high variables are located near each other

---

[2] We also examine inverse distance, which requires a weight for every block-group pair. This not only requires a larger matrix, but also had generally lower *I* statistics in our analysis.

(and low values are located near other low values):

$$G_i^* = \frac{\sum_j^n w_{ij} x_j - \bar{x} \sum_j^n w_{ij}}{s_x \sqrt{\frac{n \sum_j^n w_{ij}^2 - \left[\sum_j^n w_{ij}\right]^2}{n-1}}} \quad (3).$$

This can allow us to map the most highly-deprived, as well as the least-deprived, block groups in each city. We can also examine whether other socioeconomic variables (such as race and housing tenure) are statistically different in "hot" or "cold" spots. We do this by applying the Kruskall-Wallis (1952) test, a nonparametric analogue to ANOVA that examines the difference in means between two (or more) subgroups in a sample.[3] Here, we focus on race (the percent nonwhite, *percnonwhi*), commute (the percent who take more than 40 minutes to get to work, *perc40min*, and the percent who use public transit *percpubtra*), tenure (the percentage of units rented, *percrent*), and median income (*medinc*). If the statistic is indeed significant, we can say that the socioeconomic variables are, on average, higher or lower in deprivation "hot spots" than elsewhere.

Spearman correlations between our deprivation indices and these variables will be able to help us examine whether residents of more deprived areas are less white, poorer, or have harder commutes. We also calculate bivariate Moran's *I* statistics between each deprivation index and each socioeconomic variable, to allow for spatial relationships. Overall, we find that the six-variable measure has higher correlations with other variables than does the four-variable measure, with the variance-weighted method higher than PCA in certain cases. We also find evidence of deprivation "hot spots" that are statistically different socioeconomically than other areas. Our results are presented below.

---

[3] For a detailed explanation of various nonparametric methods, see Hollander and Wolfe (1999).

### III. Results

Using our 2014 Census data, we create our four deprivation indices for each city. In a few cases, missing data preclude the construction of one or more indices, so we omit those block groups from our analysis. One set of weights uses each component's standard deviation (included in Table 1); the PCA results for the other method are provided in Table 2. We see that there is a single principal component with an Eigenvalue above one for each city, so we use those components for our indices. In the four-variable specification, unemployment has the smallest factor loadings. Vacancy rates have slightly higher loadings, and deviations from median income have higher loadings. The income values, as expected, are negative. Maps of all indices, for each city (differentiated into five groups by each series' own natural breaks), are shown in Figure 3. They appear to be very similar visually, so the data analysis that follows will be particularly important in uncovering differences among the series.

Table 3 shows that these indices are all spatially autocorrelated, with *SD6* (the six-variable variance-weighted measure) having the largest Moran coefficient. The indices, as Table 4 indicates, are all highly correlated with one another. *PCA6* (the six-variable principal-component-based index) and *SD6* have Pearson correlations above 0.99. The Spearman and Kendall correlations are also high, although the latter values are slightly lower than the others. We infer that all of these indices might have similar explanatory power in future statistical analyses, including those that follow here.

Summary statistics for all deprivation indices are presented in Table 5. The *PCA* measures all have means of zero, while the *SD* measures are all positive; therefore, these indices might not be directly comparable with one another without some sort of transformation. The *SD* measures also have larger standard deviations than the *PCA* indices. In order to measure the dispersion of each series, we calculate Z-scores for each maximum and minimum value. Interestingly, the six-variable measures have more extreme low values, while the 4-variable measures have more "spread" at the maximum values. Most likely this is due to income's negative impact on deprivation. Many series also show evidence of nonnormality.

Next, we map Getis-Ord "hot" and "cold" spots in Table 6. In the interest of space, we do this only for our most comprehensive and mathematically sophisticated series, *PCA6*. We see deprivation "hot spots" near low-income areas: Buffalo's East Side, the East Side of Cleveland (away from suburban borders), certain neighborhoods of Detroit, and Milwaukee's Near North and Near South sides. In addition, we see relatively wealthy "cold spots," such as North Buffalo and areas near Detroit's downtown. We expect that these block groups will exhibit significant differences in socioeconomic structure when compared to other areas.

Our summary statistics, broken down by neighborhood type, and our Kruskall-Wallis results in Tables 6 and 7 confirm this hypothesis. With few exceptions, the citywide averages for our socioeconomic variables are similar to the "neutral" block groups that are neither hot nor cold. With few exceptions (such as Cleveland's nonwhite population) residents of "hot spots" are poorer and less white, and rent more and have longer commutes via public transit. The inverse is true for "cold spots." These differences are statistically significant, as shown by the Kruskal-Wallis tests.

In addition, in Table 8 we see that Spearman correlations between deprivation and our chosen socioeconomic variables are relatively high for all pairs except commute time. In general, the six-variable indices show larger correlations, with *SD6* often higher than *PCA6*. Our bivariate Moran's *I* coefficients are lower than those from the non-spatial methods, especially for Detroit. Commutes are least correlated with deprivation.

In all, we find that our four deprivation measures are broadly similar and highly correlated with one another, with a few differences. Including income and vacancies along with unemployment, poverty, education, and public assistance use tends to create indices with somewhat different statistical properties, yet that are more highly correlated with other socioeconomic variables. We can conclude that vacancy rates do indeed capture important characteristics related to urban blight, and that relative income adds to our index beyond describing poverty.

The weighting scheme also does not point toward a single preferred method; the means differ between variance-based and PCA measures, but one does not uniformly outperform the other. Focusing on one of our four indices for further statistical analysis, we find evidence of deprivation "hot spots" in each of four U.S. cities; these are shown to be poorer and less white than other areas. Detroit behaves differently than Buffalo, Cleveland, or Milwaukee, as the entire city shows characteristics that are common to smaller, confined areas in the other three. This difference manifests itself in our results.

## IV. Conclusion

Measuring "deprivation" across spatial areas has important uses in economic policy, public health, and social service provision. Yet measuring this concept appropriately is necessary if subsequent analyses are to be accurate. The choice of variables and methodology used to construct such indices has differed across the literature. Many of these studies focus on areas outside the United States (such as Britain and South Africa), with relatively few examining American inner cities.

Focusing on economic variables, and eschewing others that might be inappropriate for the U.S. or make value judgements, this study constructs four types of deprivation index for four U.S. "rust belt" cities. We find that adding two additional variables to our four-variable index captures important inner-city characteristics, and that using principal components analysis instead of variance-based weights alters the index's statistical properties somewhat, but overall the four indices are visually and statistically similar. All four indices show correlations—both nonparametric and spatial—with a set of socioeconomic variables. Using one of the four measures, we map Getis-Ord "hot spots," and use a set of statistical tests to show that these areas are indeed poorer and less white than other areas, and their residents rent more and use more public transit.

This study therefore generates two sets of key results. First, it does not appear that the choice of weighting scheme has much effect on our statistical analysis, so researchers could choose the one they prefer without jeopardizing their results. In addition, including the vacancy rate and relative index is useful when constructing inner-city deprivation indices. Second, we use these indices to imaging four large U.S. cities, finding evidence of highly deprived areas with lower quality of life and large minority populations. These indices, therefore could be used when choosing where to allocate scarce economic development or social-service resources.

**Figure 1. Median Income in Four U.S. Cities, 2014.**

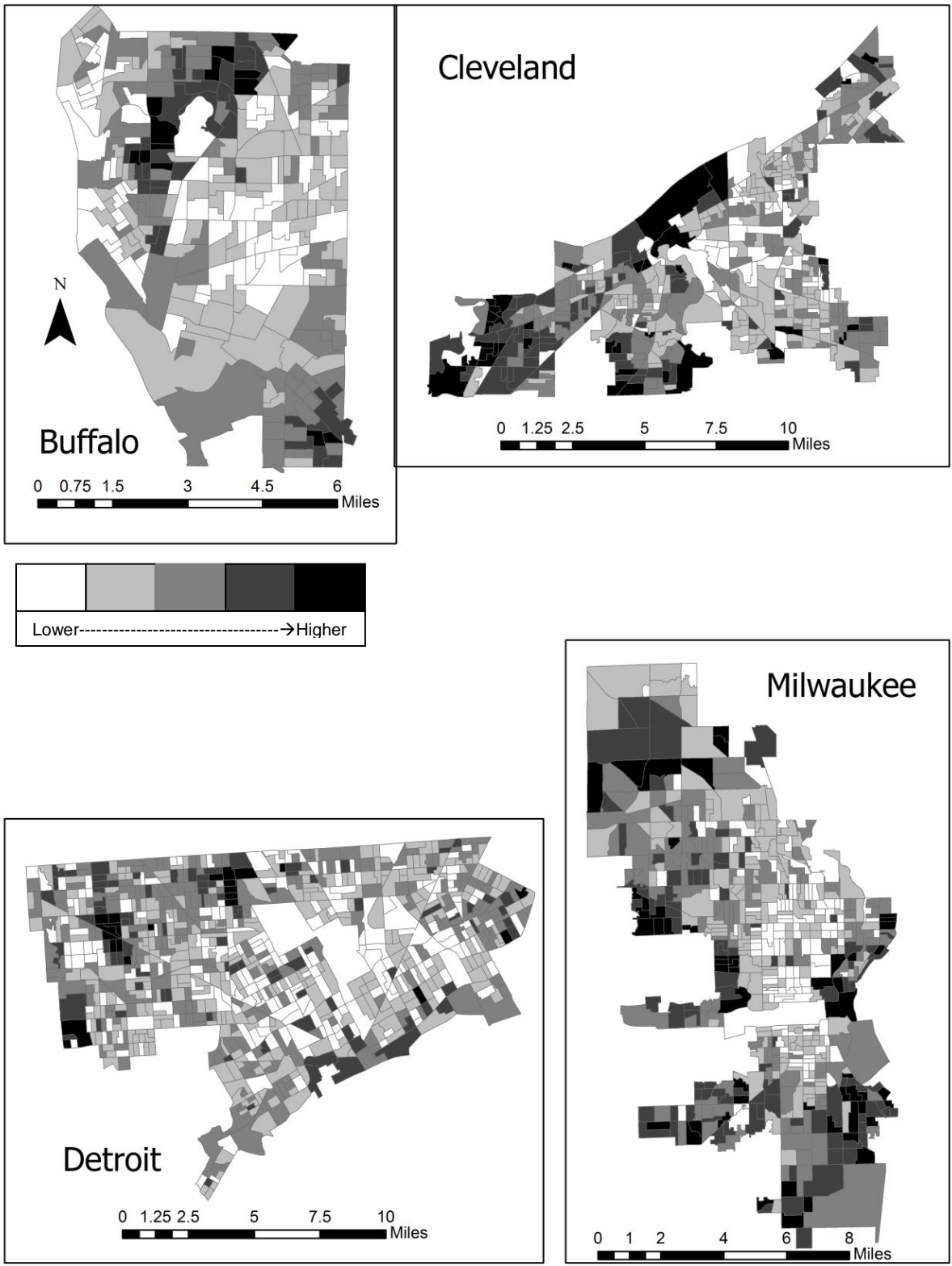

**Figure 2: Nonwhite Proportion of Population, 2014.**

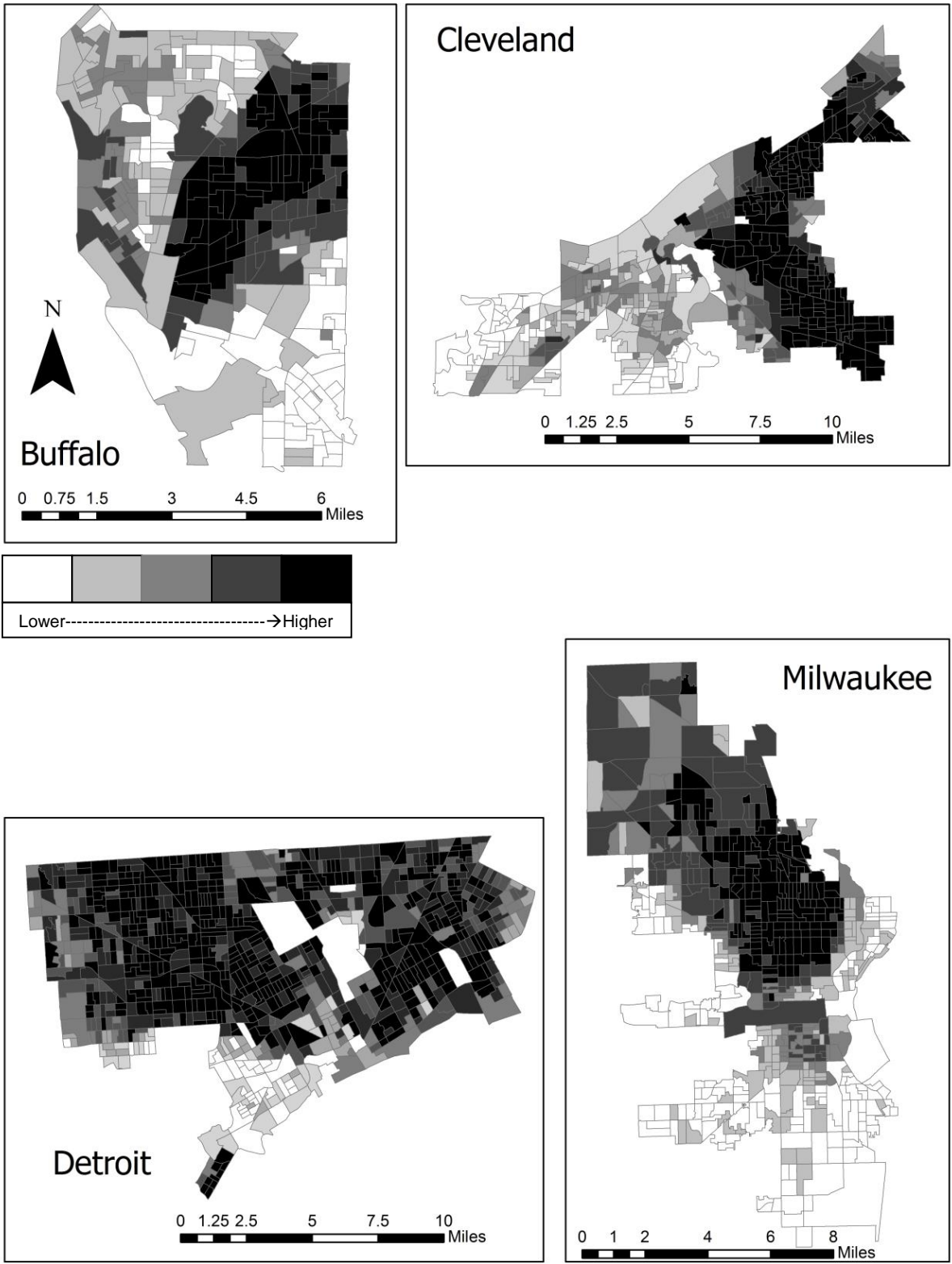

# Table 1. Summary Statistics of Index Components.

| | Buffalo | | | | | | Cleveland | | | | | |
| --- | --- | --- | --- | --- | --- | --- | --- | --- | --- | --- | --- | --- |
| | percmedy | percnohs | percpov | percsnap | percvac | unemp | percmedy | percnohs | percpov | percsnap | percvac | unemp |
| Mean | 7.63 | 18.25 | 29.62 | 34.91 | 16.91 | 7.78 | 8.65 | 23.77 | 33.57 | 36.19 | 22.28 | 11.97 |
| Median | -2.03 | 16.59 | 28.65 | 36.94 | 14.64 | 6.81 | -2.91 | 22.93 | 33.05 | 34.74 | 21.77 | 10.85 |
| Max | 238.28 | 58.55 | 85.71 | 88.93 | 58.95 | 31.53 | 270.63 | 61.84 | 93.27 | 88.45 | 55.18 | 43.67 |
| Min | -78.39 | 0 | 0 | 0 | 0 | 0 | -75.09 | 0.86 | 0 | 0 | 0 | 0 |
| SD | 53.61 | 12.08 | 16.81 | 19.84 | 11.92 | 5.58 | 53.19 | 11.97 | 17.14 | 18.84 | 12.86 | 7.86 |
| N | 283 | 283 | 283 | 283 | 283 | 283 | 457 | 457 | 457 | 457 | 457 | 457 |
| | Detroit | | | | | | Milwaukee | | | | | |
| | percmedy | percnohs | percpov | percsnap | percvac | unemp | percmedy | percnohs | percpov | percsnap | percvac | unemp |
| Mean | 6.89 | 22.92 | 37.27 | 43.39 | 30.97 | 15.09 | 5.60 | 19.53 | 27.26 | 32.25 | 11.31 | 9.01 |
| Median | -5.28 | 20.44 | 37.21 | 44.13 | 30.23 | 13.87 | -3.54 | 16.50 | 26.25 | 31.31 | 9.29 | 8.14 |
| Max | 423.27 | 76.02 | 90.77 | 100.00 | 76.42 | 84.29 | 157.65 | 75.48 | 83.93 | 87.04 | 47.27 | 34.50 |
| Min | -86.50 | 0 | 0 | 0 | 0 | 0 | -75.69 | 0 | 0 | 0 | 0 | 0 |
| SD | 53.16 | 13.47 | 16.08 | 16.82 | 15.27 | 8.79 | 47.44 | 14.31 | 17.40 | 20.56 | 9.72 | 5.97 |
| N | 871 | 871 | 871 | 871 | 871 | 871 | 567 | 567 | 567 | 567 | 567 | 567 |

# Table 2. Principal Components Analysis.

*4-Variable Specification*

| Buffalo | | | | Cleveland | | | |
| --- | --- | --- | --- | --- | --- | --- | --- |
| Eigenvalues: | | Eigenvectors (Loadings): | | Eigenvalues: | | Eigenvectors (Loadings): | |
| Number | Value | Variable | PC 1 | Number | Value | Variable | PC 1 |
| 1 | 2.492 | percnohs | 0.490 | 1 | 2.567 | percnohs | 0.446 |
| 2 | 0.821 | percpov | 0.558 | 2 | 0.727 | percpov | 0.551 |
| 3 | 0.485 | percsnap | 0.579 | 3 | 0.466 | percsnap | 0.558 |
| 4 | 0.202 | unemp | 0.338 | 4 | 0.240 | unemp | 0.433 |
| Detroit | | | | Milwaukee | | | |
| Eigenvalues: | | Eigenvectors (Loadings): | | Eigenvalues: | | Eigenvectors (Loadings): | |
| Number | Value | Variable | PC 1 | Number | value | variable | PC 1 |
| 1 | 2.017 | percnohs | 0.402 | 1 | 2.757 | percnohs | 0.480 |
| 2 | 1.006 | percpov | 0.597 | 2 | 0.642 | percpov | 0.514 |
| 3 | 0.592 | percsnap | 0.601 | 3 | 0.417 | percsnap | 0.563 |
| 4 | 0.384 | unemp | 0.347 | 4 | 0.184 | unemp | 0.435 |

*6-Variable Specification*

| Buffalo | | | | Cleveland | | | |
| --- | --- | --- | --- | --- | --- | --- | --- |
| Eigenvalues: | | Eigenvectors (Loadings): | | Eigenvalues: | | Eigenvectors (Loadings): | |
| Number | Value | Variable | PC1 | Number | Value | Variable | PC 1 |
| 1 | 3.405 | percnohs | 0.404 | 1 | 3.408 | percnohs | 0.366 |
| 2 | 0.859 | percpov | 0.473 | 2 | 0.859 | percpov | 0.480 |
| 3 | 0.786 | percsnap | 0.484 | 3 | 0.696 | percsnap | 0.475 |
| 4 | 0.501 | unemp | 0.274 | 4 | 0.528 | unemp | 0.351 |
| 5 | 0.248 | percvac | 0.273 | 5 | 0.305 | percvac | 0.269 |
| 6 | 0.200 | percmedy | -0.480 | 6 | 0.205 | percmedy | -0.462 |
| Detroit | | | | Milwaukee | | | |
| Eigenvalues: | | Eigenvectors (Loadings): | | Eigenvalues: | | Eigenvectors (Loadings): | |
| Number | Value | Variable | PC 1 | Number | Value | Variable | PC 1 |
| 1 | 2.803 | percmedy | -0.491 | 1 | 3.687 | percnohs | 0.389 |
| 2 | 1.013 | percnohs | 0.317 | 2 | 0.792 | percpov | 0.454 |
| 3 | 0.806 | percpov | 0.513 | 3 | 0.631 | percsnap | 0.480 |
| 4 | 0.645 | percsnap | 0.481 | 4 | 0.498 | unemp | 0.358 |
| 5 | 0.457 | percvac | 0.313 | 5 | 0.213 | percvac | 0.286 |
| 6 | 0.276 | unemp | 0.257 | 6 | 0.179 | percmedy | -0.449 |

**Figure 3: Alternative Measures of Economic Deprivation.**

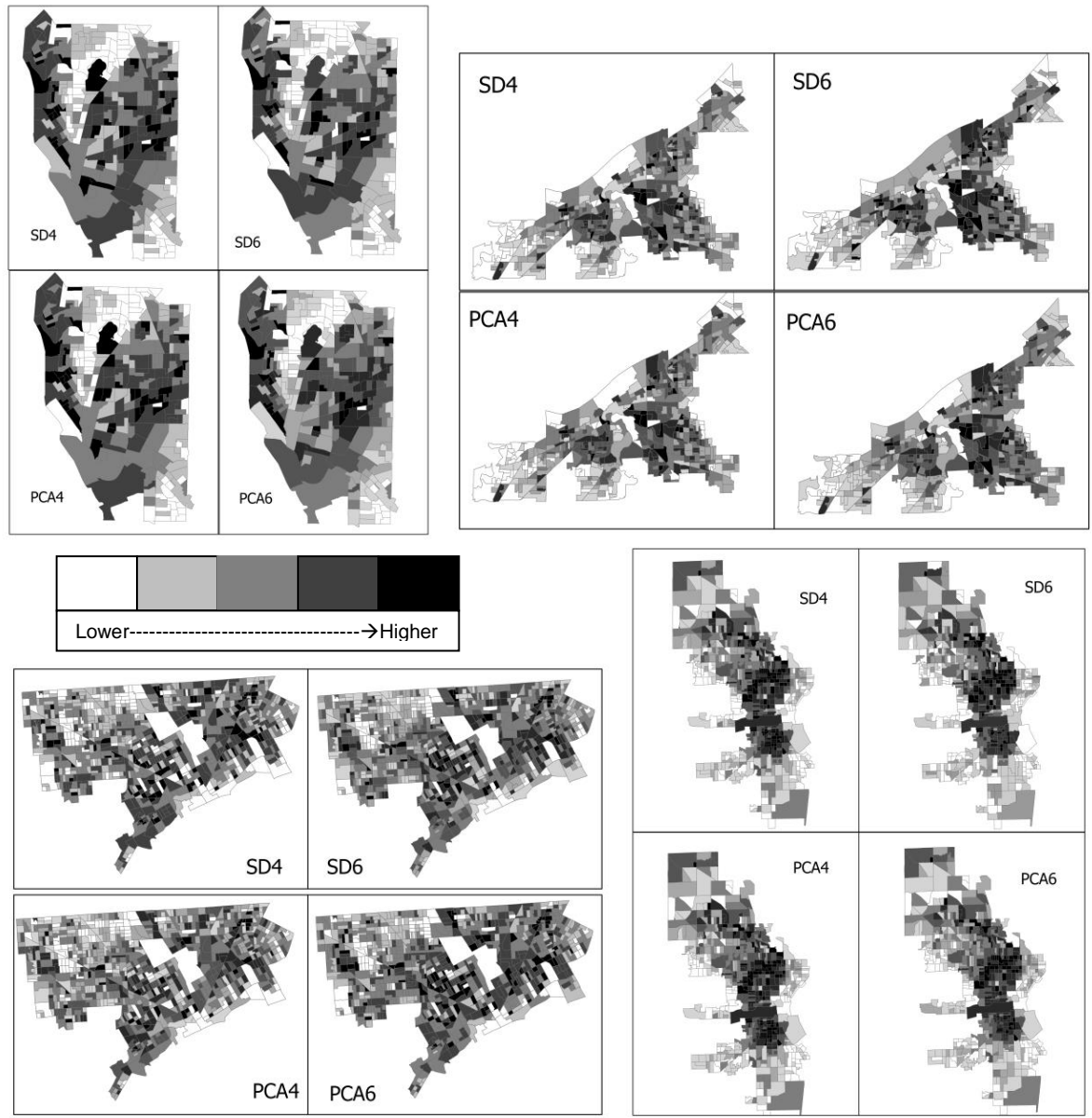

**Table 3. Moran's *I* Statistic for Spatial Autocorrelation.**

|      | Buffalo | Cleveland | Detroit | Milwaukee |
|------|---------|-----------|---------|-----------|
| SD4  | 0.504   | 0.417     | 0.279   | 0.641     |
| SD6  | 0.578   | 0.513     | 0.367   | 0.673     |
| PCA4 | 0.517   | 0.421     | 0.280   | 0.645     |
| PCA6 | 0.572   | 0.491     | 0.350   | 0.666     |

**Table 4: Correlations Among Alternative Measures of Deprivation.**

| Buffalo Pearson | SD4 | SD6 | PCA4 | Cleveland Pearson | SD4 | SD6 | PCA4 | Detroit Pearson | SD4 | SD6 | PCA4 | Milwaukee Pearson | SD4 | SD6 | PCA4 |
|---|---|---|---|---|---|---|---|---|---|---|---|---|---|---|---|
| SD4 | 1 | | | SD4 | 1 | | | SD4 | 1 | | | SD4 | 1 | | |
| SD6 | 0.963 | 1 | | SD6 | 0.962 | 1 | | SD6 | 0.951 | 1 | | SD6 | 0.970 | 1 | |
| PCA4 | 0.995 | 0.962 | 1 | PCA4 | 0.999 | 0.963 | 1 | PCA4 | 0.992 | 0.952 | 1 | PCA4 | 0.999 | 0.971 | 1 |
| PCA6 | 0.967 | 0.994 | 0.975 | PCA6 | 0.969 | 0.996 | 0.973 | PCA6 | 0.941 | 0.992 | 0.957 | PCA6 | 0.976 | 0.998 | 0.978 |
| Spearman | SD4 | SD6 | PCA4 | Spearman | SD4 | SD6 | PCA4 | Spearman | SD4 | SD6 | PCA4 | Spearman | SD4 | SD6 | PCA4 |
| SD4 | 1 | | | SD4 | 1 | | | SD4 | 1 | | | SD4 | 1 | | |
| SD6 | 0.968 | 1 | | SD6 | 0.964 | 1 | | SD6 | 0.949 | 1 | | SD6 | 0.972 | 1 | |
| PCA4 | 0.994 | 0.967 | 1 | PCA4 | 0.998 | 0.966 | 1 | PCA4 | 0.990 | 0.951 | 1 | PCA4 | 0.999 | 0.973 | 1 |
| PCA6 | 0.973 | 0.993 | 0.980 | PCA6 | 0.973 | 0.995 | 0.977 | PCA6 | 0.940 | 0.989 | 0.959 | PCA6 | 0.978 | 0.998 | 0.980 |
| Kendall | SD4 | SD6 | PCA4 | Kendall | SD4 | SD6 | PCA4 | Kendall | SD4 | SD6 | PCA4 | Kendall | SD4 | SD6 | PCA4 |
| SD4 | 1 | | | SD4 | 1 | | | SD4 | 1 | | | SD4 | 1 | | |
| SD6 | 0.847 | 1 | | SD6 | 0.836 | 1 | | SD6 | 0.809 | 1 | | SD6 | 0.851 | 1 | |
| PCA4 | 0.936 | 0.844 | 1 | PCA4 | 0.969 | 0.841 | 1 | PCA4 | 0.919 | 0.812 | 1 | PCA4 | 0.979 | 0.856 | 1 |
| PCA6 | 0.858 | 0.932 | 0.881 | PCA6 | 0.857 | 0.945 | 0.869 | PCA6 | 0.794 | 0.915 | 0.829 | PCA6 | 0.869 | 0.962 | 0.876 |

**Table 5. Summary Statistics, Indices of Deprivation.**

| | Cleveland | | | | Detroit | | | |
|---|---|---|---|---|---|---|---|---|
| | SD4 | SD6 | PCA4 | PCA6 | SD4 | SD6 | PCA4 | PCA6 |
| Mean | 7.40 | 8.97 | 0.00 | 0.00 | 8.32 | 10.21 | 0.00 | 0.00 |
| Median | 7.39 | 9.20 | 0.00 | 0.13 | 8.37 | 10.59 | 0.06 | 0.16 |
| Maximum | 17.94 | 20.77 | 5.31 | 5.13 | 20.39 | 26.09 | 5.58 | 5.88 |
| Minimum | 0.23 | -3.75 | -3.60 | -5.27 | 0.80 | -6.13 | -3.84 | -7.28 |
| Std. Dev. | 3.19 | 4.46 | 1.60 | 1.85 | 2.79 | 4.00 | 1.42 | 1.68 |
| Z(max) | 3.30 | 2.65 | 3.32 | 2.77 | 4.33 | 3.97 | 3.93 | 3.50 |
| Z(min) | -2.25 | -2.85 | -2.25 | -2.85 | -2.70 | -4.09 | -2.70 | -4.33 |
| JB (Prob) | 4.35 (0.11) | 5.59 (0.06) | 4.07 (0.13) | 4.22 (0.12) | 4.16 (0.12) | 21.08 (0.00) | 0.9 (0.64) | 35.71 (0.00) |
| N | 457 | 457 | 457 | 457 | 871 | 871 | 871 | 871 |
| | Buffalo | | | | Milwaukee | | | |
| | SD4 | SD6 | PCA4 | PCA6 | SD4 | SD6 | PCA4 | PCA6 |
| Mean | 6.43 | 7.72 | 0.00 | 0.00 | 6.01 | 7.06 | 0.00 | 0.00 |
| Median | 6.47 | 8.03 | 0.00 | 0.16 | 5.57 | 6.77 | -0.18 | -0.09 |
| Maximum | 15.53 | 21.13 | 4.09 | 4.90 | 14.52 | 18.05 | 4.22 | 4.48 |
| Minimum | 0.00 | -3.05 | -3.22 | -4.64 | 0.19 | -2.04 | -2.91 | -3.76 |
| Std. Dev. | 3.12 | 4.44 | 1.58 | 1.85 | 3.31 | 4.66 | 1.66 | 1.92 |
| JB (Prob) | 6.15 (0.05) | 4.21 (0.12) | 6.65 (0.04) | 5.11 (0.08) | 32.08 (0.00) | 25.49 (0.00) | 32.46 (0.00) | 25.64 (0.00) |
| Z(max) | 2.92 | 3.02 | 2.59 | 2.65 | 2.57 | 2.36 | 2.54 | 2.33 |
| Z(min) | -2.06 | -2.43 | -2.04 | -2.51 | -1.76 | -1.95 | -1.75 | -1.96 |
| N | 283 | 283 | 283 | 283 | 567 | 567 | 567 | 567 |

Notes: Z(max) and Z(min) =number of standard deviations that maximum and minimum values lie from series mean.
JB (prob) = Jarque-Bera normality test, with p-value in parentheses. Null hypothesis = normally distributed data.

**Figure 4. Getis-Ord "Hot" and "Cold Spots" for PCA6.**

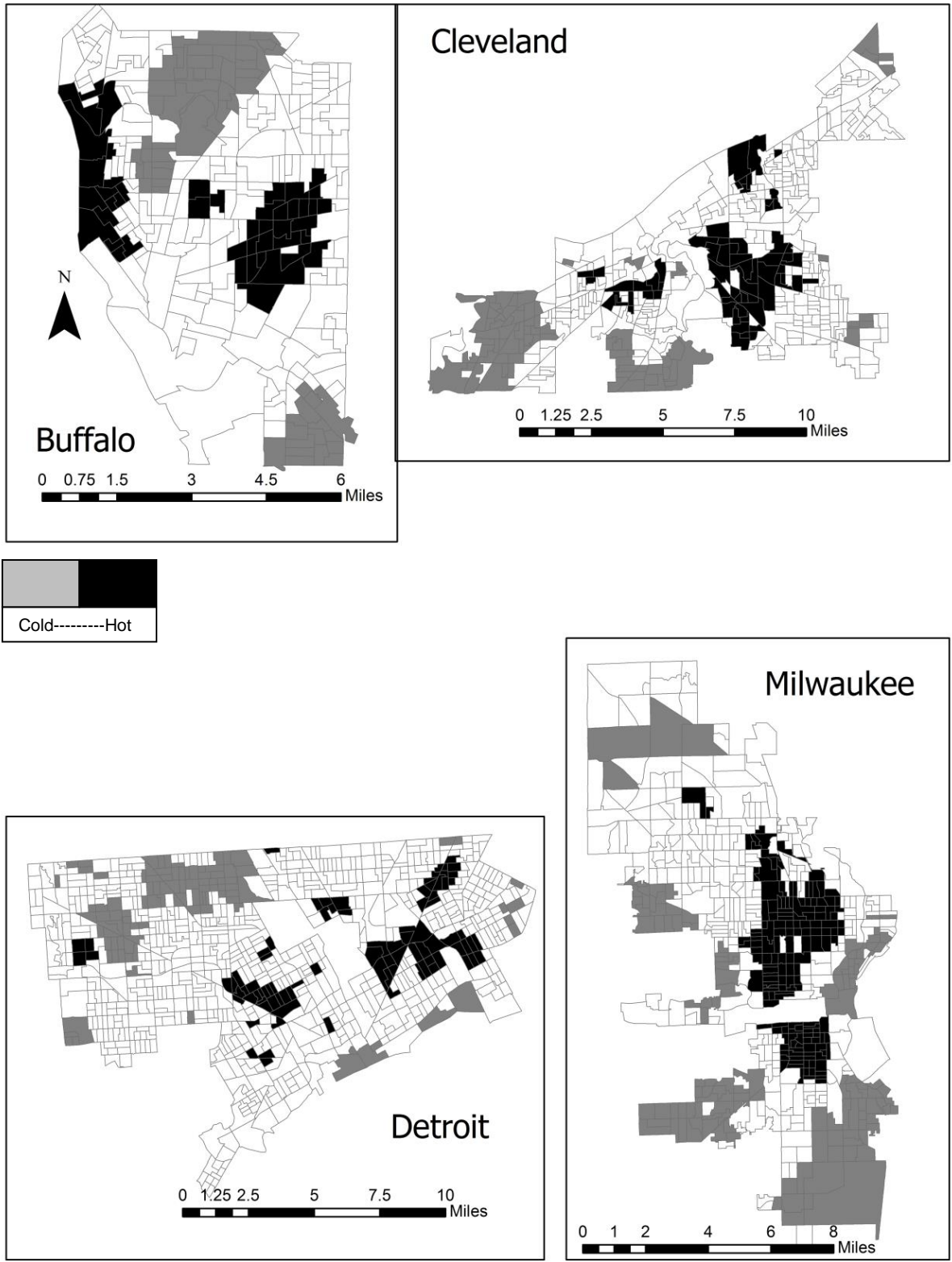

## Table 6. Summary Statistics for Socioeconomic Variables Across Groups.

| Buffalo | ALL (N=283) | | | | Hot (N=46) | | Cold (N=57) | | Neither (N=180) | |
|---|---|---|---|---|---|---|---|---|---|---|
| | Mean | SD | Min | Max | Mean | SD | Mean | SD | Mean | SD |
| perc40min | 9.8 | 9.05 | 0 | 62.6 | 14.56 | 10.92 | 5.65 | 5.29 | 9.9 | 8.89 |
| percpubtra | 13.63 | 12.96 | 0 | 62 | 21.26 | 15.23 | 4.61 | 7.59 | 14.54 | 12.11 |
| percrent | 55.76 | 21.48 | 0 | 100 | 60.14 | 17.64 | 39.45 | 19.66 | 59.81 | 20.53 |
| medinc | 34085 | 16978 | 6842 | 107125 | 21551 | 8017 | 55875 | 17438 | 30388 | 11726 |
| percnonwhi | 49.99 | 34.93 | 0 | 100 | 67.21 | 22.98 | 15.72 | 15.13 | 56.43 | 35.01 |

| Cleveland | ALL (N=457) | | | | Hot (N=65) | | Cold (N=77) | | Neither (N=315) | |
|---|---|---|---|---|---|---|---|---|---|---|
| | Mean | SD | Min | Max | Mean | SD | Mean | SD | Mean | SD |
| perc40min | 15.3 | 11.71 | 0 | 79.6 | 15.79 | 11.58 | 11.4 | 6.44 | 16.16 | 12.54 |
| percpubtra | 12.88 | 13.13 | 0 | 68 | 18.71 | 14.58 | 4.3 | 4.53 | 13.78 | 13.27 |
| percrent | 54.34 | 22.02 | 0 | 100 | 68.4 | 14.93 | 34.47 | 19.62 | 56.3 | 20.59 |
| medinc | 28442 | 13925 | 6522 | 97027 | 18081 | 5540 | 46665 | 15579 | 26126 | 10372 |
| percnonwhi | 61.01 | 35.12 | 0 | 100 | 69.01 | 28.15 | 21.87 | 23.48 | 68.92 | 32.32 |

| Detroit | ALL (N=871) | | | | Hot (N=86) | | Cold (N=101) | | Neither (N=684) | |
|---|---|---|---|---|---|---|---|---|---|---|
| | Mean | SD | Min | Max | Mean | SD | Mean | SD | Mean | SD |
| perc40min | 18.39 | 14.38 | 0 | 100 | 21.22 | 20.36 | 15.72 | 12.85 | 18.43 | 13.63 |
| percpubtra | 9.94 | 11.23 | 0 | 70 | 14.8 | 13.89 | 6.07 | 8.08 | 9.91 | 11.04 |
| percrent | 47.69 | 21.59 | 0 | 100 | 55.53 | 22.79 | 31.94 | 20.95 | 49.03 | 20.47 |
| medinc | 27892 | 13873 | 3523 | 136548 | 18312 | 7389 | 43820 | 21312 | 26744 | 11021 |
| percnonwhi | 88.38 | 19.09 | 3.9 | 100 | 93.36 | 12.55 | 92.21 | 12.18 | 87.18 | 20.4 |

| Milwaukee | ALL (N=567) | | | | Hot (N=156) | | Cold (N=128) | | Neither (N=283) | |
|---|---|---|---|---|---|---|---|---|---|---|
| | Mean | SD | Min | Max | Mean | SD | Mean | SD | Mean | SD |
| perc40min | 12.48 | 8.8 | 0 | 48 | 15.2 | 10.89 | 9.32 | 5.28 | 12.4 | 8.3 |
| percpubtra | 10.38 | 10.39 | 0 | 55 | 17.53 | 12.51 | 3.88 | 4.39 | 9.39 | 8.58 |
| percrent | 54.88 | 23.1 | 0 | 100 | 68.97 | 15.17 | 39.35 | 24.6 | 54.14 | 21.26 |
| medinc | 37477 | 16836 | 8629 | 91438 | 24028 | 8120 | 54147 | 15702 | 37350 | 13979 |
| percnonwhi | 52.18 | 35.24 | 0 | 100 | 77.25 | 23.3 | 12.16 | 13.76 | 56.47 | 31.87 |

## Table 7. Kruskall-Wallis Statistics for Differences in Means.

| Variable | Buffalo KW (p-val.) | Cleveland KW (p-val.) | Detroit KW (p-val.) | Milwaukee KW (p-val.) |
|---|---|---|---|---|
| percpubtrans | 56.85 (0.000) | 54.85 (0.000) | 24.83 (0.000) | 128.88 (0.000) |
| percrent | 38.68 (0.000) | 87.73 (0.000) | 66.73 (0.000) | 116.23 (0.000) |
| lnmedy | 113.07 (0.000) | 153.7 (0.000) | 143.02 (0.000) | 236.09 (0.000) |
| percnonwhi | 72.61 (0.000) | 113.35 (0.000) | 11.17 (0.004) | 248.83 (0.000) |
| perc40min | 19.86 (0.000) | 2.00 (0.021) | 4.60 (0.100) | 23.31 (0.000) |

**Table 8. Spearman Correlations Between Deprivation and Socioeconomic Variables.**

| Buffalo | PCA4 | PCA6 | SD4 | SD6 |
|---|---|---|---|---|
| perc40min | 0.242 | 0.257 | 0.246 | 0.265 |
| percpubtrans | 0.495 | 0.490 | 0.510 | 0.504 |
| percnonwh | 0.548 | 0.595 | 0.552 | 0.584 |
| lnMedY | -0.817 | -0.870 | -0.836 | -0.895 |
| percrent | 0.482 | 0.466 | 0.507 | 0.499 |
| *Cleveland* | PCA4 | PCA6 | SD4 | SD6 |
| perc40min | 0.185 | 0.212 | 0.190 | 0.214 |
| percpubtrans | 0.423 | 0.455 | 0.428 | 0.458 |
| percnonwh | 0.379 | 0.478 | 0.380 | 0.459 |
| lnMedY | -0.766 | -0.826 | -0.777 | -0.852 |
| percrent | 0.523 | 0.532 | 0.538 | 0.555 |
| *Detroit* | PCA4 | PCA6 | SD4 | SD6 |
| perc40min | 0.098 | 0.101 | 0.094 | 0.097 |
| percpubtrans | 0.189 | 0.236 | 0.194 | 0.239 |
| percnonwh | 0.011 | 0.059 | 0.020 | 0.064 |
| lnMedY | -0.666 | -0.761 | -0.700 | -0.812 |
| percrent | 0.368 | 0.383 | 0.389 | 0.408 |
| *Milwaukee* | PCA4 | PCA6 | SD4 | SD |
| perc40min | 0.284 | 0.289 | 0.285 | 0.287 |
| percpubtrans | 0.525 | 0.571 | 0.529 | 0.568 |
| percnonwh | 0.663 | 0.696 | 0.665 | 0.691 |
| lnMedY | -0.810 | -0.868 | -0.816 | -0.881 |
| percrent | 0.578 | 0.623 | 0.585 | 0.631 |

**Table 9. Bivariate Moran's I Between Deprivation and Socioeconomic Variables.**

| Buffalo | PCA4 | PCA6 | SD4 | SD6 |
|---|---|---|---|---|
| perc40min | 0.232 | 0.246 | 0.233 | 0.250 |
| percpubtrans | 0.322 | 0.347 | 0.322 | 0.354 |
| percnonwh | 0.407 | 0.452 | 0.433 | 0.468 |
| lnMedY | -0.454 | -0.481 | -0.453 | -0.485 |
| percrent | 0.229 | 0.226 | 0.220 | 0.219 |
| *Cleveland* | PCA4 | PCA6 | SD4 | SD6 |
| perc40min | 0.106 | 0.136 | 0.107 | 0.148 |
| percpubtrans | 0.283 | 0.273 | 0.234 | 0.279 |
| percnonwh | 0.290 | 0.381 | 0.289 | 0.400 |
| lnMedY | -0.351 | -0.393 | -0.350 | -0.402 |
| percrent | 0.260 | 0.284 | 0.254 | 0.283 |
| *Detroit* | PCA4 | PCA6 | SD4 | SD6 |
| perc40min | 0.041 | 0.047 | 0.040 | 0.045 |
| percpubtrans | 0.060 | 0.094 | 0.057 | 0.098 |
| percnonwh | -0.120 | -0.065 | -0.134 | -0.069 |
| lnMedY | -0.206 | -0.246 | -0.199 | -0.217 |
| percrent | 0.119 | 0.149 | 0.110 | 0.141 |
| *Milwaukee* | PCA4 | PCA6 | SD4 | SD6 |
| perc40min | 0.230 | 0.243 | 0.229 | 0.248 |
| percpubtrans | 0.390 | 0.422 | 0.389 | 0.429 |
| percnonwh | 0.549 | 0.580 | 0.547 | 0.586 |
| lnMedY | -0.515 | -0.543 | -0.513 | -0.545 |
| percrent | 0.354 | 0.390 | 0.351 | 0.391 |